\documentclass[12pt,a4paper,english,aps,superscriptaddress,groupedaddress,showpacs,showkeys]{revtex4}

\usepackage[T1]{fontenc}
\usepackage[latin9]{inputenc}
\usepackage{babel}
\usepackage{textcomp}
\usepackage{amsmath}
\usepackage{amssymb}
\usepackage{esint}
\usepackage[unicode=true]
 {hyperref}
\usepackage{breakurl}

\makeatletter

\@ifundefined{textcolor}{}
{%
 \definecolor{BLACK}{gray}{0}
 \definecolor{WHITE}{gray}{1}
 \definecolor{RED}{rgb}{1,0,0}
 \definecolor{GREEN}{rgb}{0,1,0}
 \definecolor{BLUE}{rgb}{0,0,1}
 \definecolor{CYAN}{cmyk}{1,0,0,0}
 \definecolor{MAGENTA}{cmyk}{0,1,0,0}
 \definecolor{YELLOW}{cmyk}{0,0,1,0}
}

\usepackage{dcolumn}
\usepackage{bm}
\usepackage{amsfonts}

\makeatother

\begin{document}

\title{On a paradox in quantum mechanics and its resolution}

\author{Padtarapan Banyadsin}

\email{[padtarapanb62@nu.ac.th]}

\affiliation{Physics Department, Naresuan University, Phitsanulok 65000, Thailand}

\author{Salvatore De Vincenzo}

\homepage{https://orcid.org/0000-0002-5009-053X}

\email{[salvatored@nu.ac.th]}

\affiliation{The Institute for Fundamental Study (IF), Naresuan University, Phitsanulok
65000, Thailand}

\date{July 6, 2023}
\begin{abstract}
\noindent Consider a free Schr\"{o}dinger particle inside an interval
with walls characterized by the Dirichlet boundary condition. Choose
a parabola as the normalized state of the particle that satisfies
this boundary condition. To calculate the variance of the Hamiltonian
in that state, one needs to calculate the mean value of the Hamiltonian
and that of its square. If one uses the standard formula to calculate
these mean values, one obtains both results without difficulty, but
the variance unexpectedly takes an imaginary value. If one uses the
same expression to calculate these mean values but first writes the
Hamiltonian and its square in terms of their respective eigenfunctions
and eigenvalues, one obtains the same result as above for the mean
value of the Hamiltonian but a different value for its square (in
fact, it is not zero); hence, the variance takes an acceptable value.
From whence do these contradictory results arise? The latter paradox
has been presented in the literature as an example of a problem that
can only be properly solved by making use of certain fundamental concepts
within the general theory of linear operators in Hilbert spaces. Here,
we carefully review those concepts and apply them in a detailed way
to resolve the paradox. Our results are formulated within the natural
framework of wave mechanics, and to avoid inconveniences that the
use of Dirac's symbolic formalism could bring, we avoid the use of
that formalism throughout the article. In addition, we obtain a resolution
of the paradox in an entirely formal way without addressing the restrictions
imposed by the domains of the operators involved. We think that the
content of this paper will be useful to undergraduate and graduate
students as well as to their instructors.
\end{abstract}



\maketitle

\section{Introduction}

\noindent As many physicists who teach relativistic and nonrelativistic
quantum mechanics already know, adequate treatment of certain quantum
systems with boundaries and/or singular potentials requires mention
of certain concepts and ideas that are specific to functional analysis,
roughly speaking, to linear algebra in vector spaces of infinite dimension.
For example, certain concepts within the theory of linear operators
in Hilbert spaces, such as unbounded operators and their domains,
are necessary. Standard books on quantum mechanics generally do not
present sufficient information on these nontrivial topics, an omission
that is understandable, but instead present a superficial analysis
using the tools of linear algebra in vector spaces of finite dimension,
i.e., the algebra of $n\times n$ matrices (apropos of this, finite-order
matrices are always bounded operators). Unfortunately, this type of
treatment can lead to paradoxes and errors in the calculations and
ultimately to false conclusions. A nice example of such a paradox
is precisely the one we consider here. Although this paradox has already
been presented and treated in various references, namely, \cite{RefA}
(pp. 1899-1900, 1924-1926), \cite{RefB} (pp. 322-323), \cite{RefC}
(pp. 168-173), \cite{RefD} (pp. 12-13, 234), \cite{RefE} (pp. 645-646),
\cite{RefF}, \cite{RefG} (see Example 2.17), we believe that it
is necessary to reanalyze it, i.e., to make a more complete and specific
study of the issues surrounding the paradox and then resolve it. That
is precisely what we do here. In addition, we do not use Dirac's symbolic
formalism in our analysis because it can lead to serious mathematical
complications that obscure or make impossible the study of unbounded
operators \cite{RefA,RefH}. 

In the remainder of this section, we present the paradox. Then, in
Section II, we carefully review and discuss the fundamental concepts
of the theory of linear operators in Hilbert spaces that are relevant
to proper analisis of the paradox. In Section III, we use the results
of Section II to resolve the paradox. A final discussion of our results
is given in Section IV. Finally, in the Appendix, we use a formal
procedure to calculate the problematic mean value that arises in the
paradox, that is, the mean value of the square of the Hamiltonian
operator for the system in question. Here, the word ``formal'' means
that we do not address the restrictions imposed by the domains of
the operators involved.

Let us consider a (free) quantum particle of mass $\mathrm{m}$ in
a one-dimensional box $\Omega=[0,L]$; thus, with Hilbert space $\mathcal{H}=\mathcal{L}^{2}(\Omega)$,
namely, the Hilbert space of square integrable functions on $\Omega$.
The self-adjoint Hamiltonian operator has the following formal action:
\begin{equation}
\hat{H}=-\frac{\hbar^{2}}{2\mathrm{m}}\frac{\mathrm{d}^{2}}{\mathrm{d}x^{2}},
\end{equation}
and its domain of definition is essentially given by
\begin{equation}
\mathcal{D}(\hat{H})=\left\{ \,\psi\mid\psi\in\mathcal{L}^{2}(\Omega)\,,\hat{H}\psi\in\mathcal{L}^{2}(\Omega)\,,\psi(0)=\psi(L)=0\,\right\} \subset\mathcal{H}.
\end{equation}
The latter Hamiltonian, with the Dirichlet boundary condition in its
domain, describes the physics of a particle that is confined to the
box. The latter boundary condition defines a box with specific impenetrable
walls; however, we know that in this problem there are an infinite
number of confining boundary conditions, i.e., infinite types of impenetrable
walls (see, for example, Ref. \cite{RefB}). As is well known, the
(real) eigenvalues $E_{N}$ and the (orthonormal) eigenfunctions $\psi_{N}(x)$
of $\hat{H}$ satisfy the relation $(\hat{H}\psi_{N})(x)=E_{N}\psi_{N}(x)$,
and are given by
\begin{equation}
E_{N}=\frac{\hbar^{2}}{2\mathrm{m}}\left(\frac{N\pi}{L}\right)^{2}\,,\quad\psi_{N}(x)=\sqrt{\frac{2}{L}}\sin\left(\frac{N\pi}{L}x\right)\,,\quad N=1,2,\ldots\,
\end{equation}
(Obviously, the eigenfunctions of $\hat{H}$ belong to $\mathcal{D}(\hat{H})$).
Certainly, when $E=0$ and $E<0$, the only eigenfunction that is
obtained is the trivial eigenfunction, i.e., $\psi_{E\leq0}(x)=0$.

Let
\begin{equation}
\psi(x)=\sqrt{\frac{30}{L^{5}}}\, x(L-x)
\end{equation}
be the normalized wavefunction of the particle, for example, at an
initial time. Note that this state is a parabola with its maximum
at $x=L/2$. Clearly, the latter $\psi(x)$ also belongs to $\mathcal{D}(\hat{H})$.
Let us calculate the root mean square deviation (or the variance)
of the Hamiltonian in the state $\psi(x)$, namely,
\begin{equation}
(\Delta\hat{H})_{\psi}=\sqrt{\langle\hat{H}^{2}\rangle_{\psi}-\langle\hat{H}\rangle_{\psi}^{2}}\,.
\end{equation}
The calculation of the mean value of $\hat{H}$ is immediate, and
the result is given by
\begin{equation}
\langle\hat{H}\rangle_{\psi}=\langle\psi,\hat{H}\psi\rangle=\int_{0}^{L}\mathrm{d}x\,\psi^{*}(x)(\hat{H}\psi)(x)=\frac{5\hbar^{2}}{\mathrm{m}L^{2}}.
\end{equation}
(The asterisk $^{*}$ denotes the complex conjugate, as usual). Alternatively,
this quantity can also be calculated using the spectral theorem, namely,
\begin{equation}
\hat{H}=\sum_{N=1}^{\infty}E_{N}\hat{P}_{N},
\end{equation}
where $\hat{P}_{N}$ is the projector operator onto the (one-dimensional)
subspace spanned by the normalized eigenvector of the Hamiltonian
$\psi_{N}$ \cite{RefI}. Thus, the action of $\hat{P}_{N}$ on a
state $\psi$ is given by $(\hat{P}_{N}\psi)(x)=\langle\psi_{N},\psi\rangle\,\psi_{N}(x)$.
Then, the mean value of $\hat{H}$ can be written as follows:
\begin{equation}
\langle\hat{H}\rangle_{\psi}=\langle\psi,\hat{H}\psi\rangle=\sum_{N=1}^{\infty}E_{N}\left|\langle\psi_{N},\psi\rangle\right|^{2},
\end{equation}
and, again, the result given in Eq. (6) is obtained, namely,
\begin{equation}
\langle\hat{H}\rangle_{\psi}=240\frac{\hbar^{2}}{\mathrm{m}L^{2}}\frac{1}{\pi^{4}}\sum_{N=1}^{\infty}\frac{(1-(-1)^{N})}{N^{4}}=240\frac{\hbar^{2}}{\mathrm{m}L^{2}}\frac{1}{\pi^{4}}\,2\frac{\pi^{4}}{96}=\frac{5\hbar^{2}}{\mathrm{m}L^{2}}.
\end{equation}
On the other hand, the calculation of the mean value of $\hat{H}^{2}$
also appears to be straightforward. Note that $(\hat{H}^{2}\psi)(x)=0$;
therefore,
\begin{equation}
\langle\hat{H}^{2}\rangle_{\psi}=\langle\psi,\hat{H}^{2}\psi\rangle=\int_{0}^{L}\mathrm{d}x\,\psi^{*}(x)(\hat{H}^{2}\psi)(x)=0.
\end{equation}
As a consequence of the results given in Eqs. (6) and (10), we obtain
that the root mean square deviation of the Hamiltonian in the state
$\psi(x)$ is an imaginary number, which is obviously impossible.
What is wrong here? Alternatively, $\langle\hat{H}^{2}\rangle_{\psi}$
could also be calculated using the spectral theorem. In this case,
from Eq. (7), we obtain
\begin{equation}
\hat{H}^{2}=\sum_{N=1}^{\infty}E_{N}^{2}\hat{P}_{N}.
\end{equation}
However, now, we obtain a value different from zero for the mean value
of $\hat{H}^{2}$, namely,
\[
\langle\hat{H}^{2}\rangle_{\psi}=\langle\psi,\hat{H}^{2}\psi\rangle=\langle\psi,\sum_{N=1}^{\infty}E_{N}^{2}\hat{P}_{N}\,\psi\rangle=\sum_{N=1}^{\infty}E_{N}^{2}\langle\psi_{N},\psi\rangle\langle\psi,\psi_{N}\rangle=\sum_{N=1}^{\infty}E_{N}^{2}\left|\langle\psi_{N},\psi\rangle\right|^{2}
\]
\begin{equation}
=120\frac{\hbar^{4}}{\mathrm{m^{2}}L^{4}}\frac{1}{\pi^{2}}\sum_{N=1}^{\infty}\frac{(1-(-1)^{N})}{N^{2}}=120\frac{\hbar^{4}}{\mathrm{m^{2}}L^{4}}\frac{1}{\pi^{2}}\,2\frac{\pi^{2}}{8}=\frac{30\hbar^{4}}{\mathrm{m}^{2}L^{4}},
\end{equation}
and we can report the following value for the root mean square deviation
of the Hamiltonian in the state $\psi(x)$:
\begin{equation}
(\Delta\hat{H})_{\psi}=\frac{\sqrt{5}\hbar^{2}}{\mathrm{m}L^{2}}.
\end{equation}
Clearly, the result in Eq. (12) conflicts with the result given in
Eq. (10). Which of the two results is correct? What is the source
of this inconsistency? As we will see below, adequate answers to these
questions can only be obtained through judicious use of the mathematical
formalism of quantum mechanics.

\section{Preliminaries}

\noindent Some of the most important operators found in quantum mechanics
are unbounded operators. An unbounded operator is characterized by
(a) its formal action, which is the way the operator acts, and (b)
its domain, which is the subspace of the Hilbert space on which the
operator can act. For example, the differential operator $\hat{H}$
in Eq. (1) is an example of an unbounded operator. 

Let $\hat{A}$ be an unbounded linear operator from $\mathcal{H}$
into $\mathcal{H}$ ($\mathcal{H}$ is the Hilbert space). We define
a domain for $\hat{A}$, $\mathcal{D}(\hat{A})$ to ensure that $\hat{A}\chi\in\mathcal{H}$
for $\chi\in\mathcal{D}(\hat{A})\subset\mathcal{H}$. Let $\hat{B}$
be another unbounded linear operator; by multiplying $\hat{A}$ by
$\hat{B}$, $\hat{A}\hat{B}$, we have that $(\hat{A}\hat{B})\chi=\hat{A}(\hat{B}\chi)$,
where $\chi\in\mathcal{D}(\hat{A}\hat{B})$, i.e., $\mathcal{D}(\hat{A}\hat{B})=\{\,\chi\mid\chi\in\mathcal{D}(\hat{B})\;\mathrm{and}\;\hat{B}\chi\in\mathcal{D}(\hat{A})\}$.
Thus, in general, $\mathcal{D}(\hat{A}\hat{B})\neq\mathcal{D}(\hat{B}\hat{A})$,
and therefore, $\hat{A}\hat{B}\neq\hat{B}\hat{A}$; and even $\hat{A}^{2}\neq\hat{A}\hat{A}$,
because $\mathcal{D}(\hat{A}^{2})$ and $\mathcal{D}(\hat{A}\hat{A})$
may not coincide. Remember that two operators $\hat{A}$ and $\hat{B}$
are equal if their actions are equal, but their domains must also
be equal. If this is the case, one writes $\hat{A}=\hat{B}$. For
example, let us consider the operator $\hat{H}^{2}=\hat{H}\hat{H}$,
where the action of $\hat{H}$ is given in Eq. (1) and its domain
is given in Eq. (2). Clearly, the formal action of this operator is
given by 
\begin{equation}
\hat{H}^{2}=\hat{H}\hat{H}=\frac{\hbar^{4}}{4\mathrm{m^{2}}}\frac{\mathrm{d}^{4}}{\mathrm{d}x^{4}}.
\end{equation}
We have that $\hat{H}^{2}\psi=(\hat{H}\hat{H})\psi=\hat{H}(\hat{H}\psi)$,
where $\psi\in\mathcal{D}(\hat{H}^{2})=\mathcal{D}(\hat{H}\hat{H})$.
Thus, the domain of definition of this operator is given by 
\begin{equation}
\mathcal{D}(\hat{H}^{2})=\{\,\psi\mid\psi\in\mathcal{D}(\hat{H})\;\mathrm{and}\;\hat{H}\psi\in\mathcal{D}(\hat{H})\}.
\end{equation}
Note that the functions $\psi$ belonging to $\mathcal{D}(\hat{H}^{2})$
must satisfy the Dirichlet boundary condition, namely, $\psi(0)=\psi(L)=0$
because $\psi\in\mathcal{D}(\hat{H})$. Similarly, because $\hat{H}\psi\in\mathcal{D}(\hat{H})$,
the second derivative of $\psi$ must also satisfy the Dirichlet boundary
condition, namely, $\psi''(0)=\psi''(L)=0$ (henceforth, we use prime
notation to represent spatial derivatives, as usual). Then, the domain
of the square of the Hamiltonian operator given in Eq. (1), namely,
$\hat{H}^{2}=\hat{H}\hat{H}$, can be essentially written as follows:
\begin{equation}
\mathcal{D}(\hat{H}^{2})=\left\{ \,\psi\mid\psi\,,\hat{H}\psi\,,\hat{H}^{2}\psi\in\mathcal{L}^{2}(\Omega)\,,\psi(0)=\psi(L)=0\;\mathrm{and}\;\psi''(0)=\psi''(L)=0\,\right\} \subset\mathcal{H}.
\end{equation}
Clearly, the state $\psi(x)$ given in Eq. (4) does not belong to
$\mathcal{D}(\hat{H}^{2})$.

The adjoint operator of $\hat{A}$ (or its Hermitian conjugate), $\hat{A}^{\dagger}$,
is defined as follows: we say that there is an operator $\hat{A}^{\dagger}$
that satisfies
\begin{equation}
\langle\hat{A}\chi,\varphi\rangle=\langle\chi,\hat{A}^{\dagger}\varphi\rangle,
\end{equation}
where $\chi\in\mathcal{D}(\hat{A})$ and $\varphi\in\mathcal{D}(\hat{A}^{\dagger})$.
In general, the form in which $\hat{A}^{\dagger}$ acts can be determined
from Eq. (17), and its domain can be written as follows: 
\begin{equation}
\mathcal{D}(\hat{A}^{\dagger})=\left\{ \,\varphi\mid\varphi\in\mathcal{H}\mid\exists\,\xi=\hat{A}^{\dagger}\varphi\in\mathcal{\mathcal{H}}\mid\langle\hat{A}\chi,\varphi\rangle=\langle\chi,\xi\rangle\,,\forall\chi\in\mathcal{D}(\hat{A})\,\right\} .
\end{equation}
For example, applying the method of integration by parts twice, it
can be shown that the Hamiltonian $\hat{H}$ given in Eq. (1) satisfies
the following relation:
\[
\langle\hat{H}\psi,\phi\rangle=\int_{0}^{L}\mathrm{d}x\,(\hat{H}\psi)^{*}(x)\,\phi(x)=\frac{\hbar^{2}}{2\mathrm{m}}\left.\left[\,\psi^{*}(x)\,\phi'(x)-\psi'^{*}(x)\,\phi(x)\,\right]\right|_{0}^{L}
\]
\begin{equation}
+\int_{0}^{L}\mathrm{d}x\,\psi^{*}(x)\,\left(-\frac{\hbar^{2}}{2\mathrm{m}}\frac{\mathrm{d}^{2}}{\mathrm{d}x^{2}}\right)\phi(x),
\end{equation}
where $\left.\left[\, f\,\right]\right|_{0}^{L}\equiv f(L)-f(0)$.
Then, imposing the boundary condition on $\psi\in\mathcal{D}(\hat{H})$,
and identifying the action of $\hat{H}^{\dagger}$ (which is clearly
the same as that of $\hat{H}$), we can write
\begin{equation}
\langle\hat{H}\psi,\phi\rangle=\frac{\hbar^{2}}{2\mathrm{m}}\left[\,-\psi'^{*}(L)\,\phi(L)+\psi'^{*}(0)\,\phi(0)\,\right]+\langle\psi,\hat{H}^{\dagger}\phi\rangle.
\end{equation}
Evidently, the cancellation of the boundary term in Eq. (20) only
depends on imposing the Dirichlet boundary condition on $\phi\in\mathcal{D}(\hat{H}^{\dagger})$,
namely, $\phi(0)=\phi(L)=0$ (the latter would be the so-called adjoint
boundary condition). Thus, in conclusion, 
\begin{equation}
\hat{H}^{\dagger}=-\frac{\hbar^{2}}{2\mathrm{m}}\frac{\mathrm{d}^{2}}{\mathrm{d}x^{2}},
\end{equation}
and its domain of definition is given by
\begin{equation}
\mathcal{D}(\hat{H}^{\dagger})=\left\{ \,\phi\mid\phi\in\mathcal{L}^{2}(\Omega)\,,\hat{H}^{\dagger}\phi\in\mathcal{L}^{2}(\Omega)\,,\phi(0)=\phi(L)=0\,\right\} .
\end{equation}
Because the actions of $\hat{H}$ and \textrm{$\hat{H}^{\dagger}$}
are equal and their domains are also equal, we can say that $\hat{H}$
and \textrm{$\hat{H}^{\dagger}$} are equal, namely, $\hat{H}=\hat{H}^{\dagger}$. 

We can repeat the latter procedure to obtain the adjoint of the operator
$\hat{H}^{2}$ with its domain given in Eq. (16). Indeed, by applying
the method of integration by parts four times, we can see that $\hat{H}^{2}$
given in Eq. (14) verifies the following relation: 
\[
\langle\hat{H}^{2}\psi,\phi\rangle=\int_{0}^{L}\mathrm{d}x\,(\hat{H}^{2}\psi)^{*}(x)\,\phi(x)
\]
\[
=\frac{\hbar^{4}}{4\mathrm{m^{2}}}\left.\left[\,\psi'''^{*}(x)\,\phi(x)-\psi''^{*}(x)\,\phi'(x)+\psi'^{*}(x)\,\phi''(x)-\psi^{*}(x)\,\phi'''(x)\,\right]\right|_{0}^{L}
\]
\begin{equation}
+\int_{0}^{L}\mathrm{d}x\,\psi^{*}(x)\,\left(\frac{\hbar^{4}}{4\mathrm{m^{2}}}\frac{\mathrm{d}^{4}}{\mathrm{d}x^{4}}\right)\phi(x).
\end{equation}
Similarly, using the boundary conditions on $\psi\in\mathcal{D}(\hat{H}^{2})$,
and identifying the action of $(\hat{H}^{2})^{\dagger}$ (which is
the same as that of $\hat{H}^{2}$), we can write 
\[
\langle\hat{H}^{2}\psi,\phi\rangle=\frac{\hbar^{4}}{4\mathrm{m^{2}}}\left[\,\psi'''^{*}(L)\,\phi(L)+\psi'^{*}(L)\,\phi''(L)-\psi'''^{*}(0)\,\phi(0)-\psi'^{*}(0)\,\phi''(0)\,\right]
\]
\begin{equation}
+\langle\psi,(\hat{H}^{2})^{\dagger}\phi\rangle.
\end{equation}
The cancellation of the boundary term in the latter relation only
depends on imposing the Dirichlet boundary conditions on $\phi$ and
$\phi''$, namely, $\phi(0)=\phi(L)=0$ and $\phi''(0)=\phi''(L)=0$.
Thus, the adjoint of $\hat{H}^{2}$ is given by
\begin{equation}
(\hat{H}^{2})^{\dagger}=\frac{\hbar^{4}}{4\mathrm{m^{2}}}\frac{\mathrm{d}^{4}}{\mathrm{d}x^{4}},
\end{equation}
and its domain is essentially given by
\begin{equation}
\mathcal{D}((\hat{H}^{2})^{\dagger})=\left\{ \,\phi\mid\phi\,,\hat{H}^{\dagger}\phi\,,(\hat{H}^{2})^{\dagger}\phi\in\mathcal{L}^{2}(\Omega)\,,\phi(0)=\phi(L)=0\;\mathrm{and}\;\phi''(0)=\phi''(L)=0\,\right\} .
\end{equation}
Clearly, the actions of $\hat{H}^{2}$ and $(\hat{H}^{2})^{\dagger}$
are equal, and their domains are equal; thus, we can write the relation
$\hat{H}^{2}=(\hat{H}^{2})^{\dagger}$. Certainly, $(\hat{H}^{2})^{\dagger}=\hat{H}^{\dagger}\hat{H}^{\dagger}=(\hat{H}^{\dagger})^{2}$.

The definition of the adjoint given above requires that the operator
$\hat{A}$ defined on $\mathcal{H}$ be densely defined, i.e., that
its domain $\mathcal{D}(\hat{A})$ be dense in $\mathcal{H}$ (and
indeed we assume it here). Roughly speaking, this means that for any
$\chi\in\mathcal{H}$ there is a sequence of elements of $\mathcal{D}(\hat{A})$
that converges to $\chi$. In other words, any element of $\mathcal{H}$
can be obtained as a limit of functions in $\mathcal{D}(\hat{A})$.
If an unbounded operator is densely defined, then its adjoint is unique.
In fact, for a given $\varphi\in\mathcal{\mathcal{H}}$, $\xi=\hat{A}^{\dagger}\varphi$
is unique (see the definition of $\hat{A}^{\dagger}$ in Eq. (18)).
For a simple example of an operator whose domain is not dense in the
underlying Hilbert space, see Example 2.12 in Ref. \cite{RefG}. As
we know, the operators or observables in quantum mechanics must be
self-adjoint operators; hence, the eigenvalues of these operators
are real, and their eigenfunctions are orthogonal. In particular,
to generate a unitary time transformation, the Hamiltonian operator
must be self-adjoint. An operator $\hat{A}$ is self-adjoint if the
equality $\hat{A}=\hat{A}^{\dagger}$ is verified, i.e., if the actions
of $\hat{A}$ and $\hat{A}^{\dagger}$ and their domains are equal.
An operator is only Hermitian (or symmetric for mathematicians) if
the actions of $\hat{A}$ and $\hat{A}^{\dagger}$ are the same (on
the domain $\mathcal{D}(\hat{A})$), but $\mathcal{D}(\hat{A})\subset\mathcal{D}(\hat{A}^{\dagger})$,
and one writes $\hat{A}\subset\hat{A}^{\dagger}$. Thus, not all Hermitian
operators are self-adjoint, but any self-adjoint operator is Hermitian.
Clearly, the operators $\hat{H}$ and its square $\hat{H}^{2}=\hat{H}\hat{H}$
given above are both self-adjoint operators. Incidentally, if an operator
is self-adjoint, its square is also a self-adjoint operator (see Ref.
\cite{RefJ}, p. 32). 

Let us consider a set of vectors $\left\{ \chi_{N}(x)\right\} $ ($N=1,2,\ldots$)
that composes a discrete orthonormal basis of the corresponding Hilbert
space (i.e., $\langle\chi_{N},\chi_{M}\rangle=\delta_{N,M}$, where
$\delta_{N,M}$ is the Kronecker symbol, as usual). The so-called
projection operator $\hat{P}_{N}$ onto the normalized state $\chi_{N}$
has the following action on a state\textbf{ }$\chi$:\textbf{
\begin{equation}
(\hat{P}_{N}\chi)(x)=\langle\chi_{N},\chi\rangle\,\chi_{N}(x).
\end{equation}
}This is a bounded operator; thus, its domain is the whole Hilbert
space (i.e., $\chi\in\mathcal{H}$). Essentially, two properties define
this operator as a projector. The first of these properties is (a)
$\hat{P}_{N}$ is self-adjoint. Thus, we have
\[
\langle\hat{P}_{N}\chi,\varphi\rangle=\left\langle \langle\chi_{N},\chi\rangle\chi_{N},\varphi\right\rangle =\langle\chi_{N},\chi\rangle^{*}\langle\chi_{N},\varphi\rangle=\langle\chi_{N},\varphi\rangle\langle\chi,\chi_{N}\rangle=\left\langle \chi,\langle\chi_{N},\varphi\rangle\chi_{N}\right\rangle 
\]
\begin{equation}
=\langle\chi,\hat{P}_{N}\varphi\rangle,
\end{equation}
but according to the definition of the adjoint operator given in Eq.
(17), the right-hand side of the latter relation must be equal to
$\langle\chi,\hat{P}_{N}^{\dagger}\varphi\rangle$; thus, we have
that $\hat{P}_{N}$ is a self-adjoint operator, i.e., $\hat{P}_{N}=\hat{P}_{N}^{\dagger}$.
The other important property is (b) $\hat{P}_{N}^{2}=\hat{P}_{N}$.
Thus, we have
\[
\hat{P}_{N}\hat{P}_{M}\,\chi(x)=\hat{P}_{N}(\hat{P}_{M}\chi)(x)=\hat{P}_{N}\langle\chi_{M},\chi\rangle\,\chi_{M}(x)=\langle\chi_{M},\chi\rangle(\hat{P}_{N}\chi_{M})(x)
\]
\[
=\langle\chi_{M},\chi\rangle\langle\chi_{N},\chi_{M}\rangle\,\chi_{N}(x)=\langle\chi_{M},\chi\rangle\,\delta_{N,M}\,\chi_{N}(x)
\]
\[
=\left\{ \begin{array}{c}
\langle\chi_{N},\chi\rangle\,\chi_{N}(x)\\
0
\end{array}\right.\left.\begin{array}{c}
,\quad{\scriptstyle M=N}\\
,\quad{\scriptstyle M\neq N}
\end{array}\right.
\]
\[
=\left\{ \begin{array}{c}
(\hat{P}_{N}\chi)(x)\\
0
\end{array}\right.\left.\begin{array}{c}
,\quad{\scriptstyle M=N}\\
,\quad{\scriptstyle M\neq N}
\end{array}\right.
\]
\[
=\delta_{N,M}\,(\hat{P}_{N}\chi)(x)=\delta_{N,M}\,\hat{P}_{N}\,\chi(x).
\]
Therefore, 
\begin{equation}
\hat{P}_{N}\hat{P}_{M}=\delta_{N,M}\,\hat{P}_{N}\quad\Rightarrow\quad\hat{P}_{N}^{2}=\hat{P}_{N}.
\end{equation}
More precisely, the bounded operator $\hat{P}_{N}$ is a projection
operator (if and only if) because it is a self-adjoint operator and
because its square is the same operator \cite{RefI}. Additionally,
the formula $\hat{P}_{N}\hat{P}_{M}=\delta_{N,M}\,\hat{P}_{N}$ in
Eq. (29) indicates that the projectors $\hat{P}_{N}$ and $\hat{P}_{M}$
are mutually orthogonal. Incidentally, an important result of operator
theory tells us that any self-adjoint operator is bounded if and only
if its spectrum is bounded \cite{RefI}. For example, let us write
the eigenvalue equation for $\hat{P}_{N}$, that is, $(\hat{P}_{N}\,\eta)(x)=\lambda\,\eta(x)$,
where $\lambda$ is an eigenvalue and $\eta(x)$ is an eigenvector
corresponding to that eigenvalue. Clearly, we have that $(\hat{P}_{N}\hat{P}_{N}\,\eta)(x)=\lambda\,(\hat{P}_{N}\,\eta)(x)=\lambda^{2}\eta(x)$.
Using the property $\hat{P}_{N}^{2}=\hat{P}_{N}$ in the latter expression,
we obtain $\lambda\eta(x)=\lambda^{2}\eta(x)$, and therefore $\lambda=1$
and $\lambda=0$. Note that the eigenvalue equation for $\hat{P}_{N}$
is given by $\langle\chi_{N},\eta\rangle\,\chi_{N}(x)=\lambda\,\eta(x)$.
Clearly, $\eta(x)=\chi_{N}(x)$ is the eigenvector of $\hat{P}_{N}$
with the (simple) eigenvalue $\lambda=1$; also, all the functions
$\eta(x)=\chi_{M}(x)$ with $M\neq N$ are eigenvectors of $\hat{P}_{N}$
with eigenvalue $\lambda=0$, i.e., the latter eigenvalue is infinitely
degenerate. In conclusion, the spectrum of $\hat{P}_{N}=\hat{P}_{N}^{\dagger}$
is bounded, and therefore $\hat{P}_{N}$ is a bounded operator. 

However, because any function $\chi$ of $\mathcal{H}$ can be expanded
in terms of the elements of a basis, e.g., the basis formed by the
functions $\left\{ \chi_{N}(x)\right\} $, it follows that
\begin{equation}
\chi(x)=\sum_{N=1}^{\infty}\langle\chi_{N},\chi\rangle\,\chi_{N}(x)=\sum_{N=1}^{\infty}(\hat{P}_{N}\chi)(x)=\left(\sum_{N=1}^{\infty}\hat{P}_{N}\right)\chi(x)\quad\Rightarrow\quad\sum_{N=1}^{\infty}\hat{P}_{N}=\hat{1},
\end{equation}
where $\hat{1}$ is the identity operator. The latter equality expresses
the completeness of the eigenstates. Alternatively, we can write this
property as follows: 
\[
\chi(x)=\sum_{N=1}^{\infty}\langle\chi_{N},\chi\rangle\,\chi_{N}(x)=\sum_{N=1}^{\infty}\left[\int_{0}^{L}\mathrm{d}y\,\chi_{N}^{*}(y)\,\chi(y)\right]\chi_{N}(x)
\]
\begin{equation}
=\int_{0}^{L}\mathrm{d}y\left[\sum_{N=1}^{\infty}\chi_{N}(x)\chi_{N}^{*}(y)\right]\chi(y)\quad\Rightarrow\quad\sum_{N=1}^{\infty}\chi_{N}(x)\chi_{N}^{*}(y)=\delta(x-y),
\end{equation}
where $\delta(x-y)$ is the Dirac delta function, as usual (recall
the definition of the Dirac delta distribution, which for physicists
is essentially the so-called sifting property of the Dirac delta \cite{RefK}).
The latter infinite sum is the closure relation and can be interpreted
as a distribution (or a generalized function), namely, the Dirac delta
distribution \cite{RefL}. 

Let us assume that the spectrum of an unbounded self-adjoint operator
$\hat{A}$ is entirely discrete, i.e., the set of its eigenvalues
is a discrete sequence of (infinite) values in which each eigenvalue
is nondegenerate. The so-called spectral decomposition of $\hat{A}$
is written as follows: 
\begin{equation}
\hat{A}=\sum_{N=1}^{\infty}\lambda_{N}\hat{P}_{N},
\end{equation}
where $\lambda_{N}$ are the eigenvalues of $\hat{A}$ and the action
of the projector is given by $(\hat{P}_{N}\:\,\cdot\:)(x)=\langle\chi_{N},\:\cdot\:\,\rangle\,\chi_{N}(x)$,
where $\chi_{N}$ are the (normalized) eigenvectors of $\hat{A}$
(certainly, the point enclosed between parentheses and angle brackets
represents the function on which $\hat{P}_{N}$ expects to act). Thus,
$\hat{A}$ can be written in terms of its eigenvalues and eigenvectors.
This result is known as the spectral theorem (in fact, Eq. (32) is
only one of various ways in which this theorem is presented) and is
one of the most important results of operator theory (incidentally,
it is only valid for self-adjoint operators) \cite{RefA,RefE,RefI}.
It is worth mentioning that the formal infinite sum in Eq. (32) is
generally divergent but that it can be regulated and evaluated so
that it makes sense. The latter is also valid for the infinite series
given in Eq. (31). For a nice discussion of this issue, see Ref. \cite{RefM}.
Consistently, the spectral theorem implies the corresponding equation
for the eigenvalues, namely, 
\[
\hat{A}\,\chi_{N}(x)=(\hat{A}\chi_{N})(x)=\sum_{M=1}^{\infty}\lambda_{M}(\hat{P}_{M}\chi_{N})(x)=\sum_{M=1}^{\infty}\lambda_{M}\langle\chi_{M},\chi_{N}\rangle\,\chi_{M}(x)
\]
\begin{equation}
=\sum_{M=1}^{\infty}\lambda_{M}\,\delta_{M,N}\,\chi_{M}(x)=\lambda_{N}\,\chi_{N}(x)
\end{equation}
(in the last step, we also use the orthogonality of the eigenfunctions
$\chi_{N}$). The square of the operator $\hat{A}$ can be obtained
immediately from the spectral decomposition of $\hat{A}$ and through
the use of the orthogonality property of the projectors given in Eq.
(29), namely, 
\[
\hat{A}^{2}=\hat{A}\hat{A}=\left(\sum_{N=1}^{\infty}\lambda_{N}\hat{P}_{N}\right)\left(\sum_{M=1}^{\infty}\lambda_{M}\hat{P}_{M}\right)=\sum_{N=1}^{\infty}\sum_{M=1}^{\infty}\lambda_{N}\lambda_{M}\hat{P}_{N}\hat{P}_{M}
\]
\begin{equation}
=\sum_{N=1}^{\infty}\sum_{M=1}^{\infty}\lambda_{N}\lambda_{M}\,\delta_{N,M}\,\hat{P}_{N}=\sum_{N=1}^{\infty}\lambda_{N}^{2}\hat{P}_{N}.
\end{equation}
Certainly, from the latter relation, we obtain the eigenvalue equation
for $\hat{A}^{2}$, namely, $\hat{A}^{2}\,\chi_{N}(x)=\lambda_{N}^{2}\,\chi_{N}(x)$
(the procedure is essentially the same as that which led us to the
eigenvalue equation for $\hat{A}$ in Eq. (33)). 

Finally, the mean value of operator $\hat{A}$ in the (normalized)
state $\chi$ is, by definition, given by
\begin{equation}
\langle\hat{A}\rangle_{\chi}\equiv\sum_{N=1}^{\infty}\lambda_{N}\,\mathcal{P}(\lambda_{N})=\sum_{N=1}^{\infty}\lambda_{N}\left|\langle\chi_{N},\chi\rangle\right|^{2},
\end{equation}
where $\lambda_{N}$ are the eigenvalues of $\hat{A}$ and $\chi_{N}$
are its eigenvectors. According to one of the postulates of quantum
mechanics, the quantity $\mathcal{P}(\lambda_{N})=\left|\langle\chi_{N},\chi\rangle\right|^{2}$
is the probability of finding the (nondegenerate) eigenvalue $\lambda_{N}$
when the physical quantity associated with the operator $\hat{A}$
is measured on a system that is in the normalized state $\chi$ \cite{RefH}.
Certainly, the definition given in Eq. (35) is the most natural definition
that can be given for the average value of an operator. 

\section{Resolution of the paradox}

\noindent Let us consider the Hamiltonian operator given in Eqs. (1)
and (2), with its eigenfunctions and eigenvalues given in Eq. (3).
The normalized state of the particle $\psi$ is given in Eq. (4).
From Eq. (35), we can write the mean value of $\hat{H}$, namely,
\begin{equation}
\langle\hat{H}\rangle_{\psi}=\sum_{N=1}^{\infty}E_{N}\left|\langle\psi_{N},\psi\rangle\right|^{2}.
\end{equation}
Developing the latter definition further, we can write the following
expressions:
\[
\langle\hat{H}\rangle_{\psi}=\sum_{N=1}^{\infty}E_{N}\langle\psi_{N},\psi\rangle\langle\psi,\psi_{N}\rangle=\sum_{N=1}^{\infty}\langle E_{N}\psi_{N},\psi\rangle\langle\psi,\psi_{N}\rangle=\sum_{N=1}^{\infty}\langle\hat{H}\psi_{N},\psi\rangle\langle\psi,\psi_{N}\rangle,
\]
the latter is true because the eigenvalues $E_{N}$ of $\hat{H}$
are real. Thus, 
\[
\langle\hat{H}\rangle_{\psi}=\sum_{N=1}^{\infty}\langle\psi_{N},\hat{H}\psi\rangle\langle\psi,\psi_{N}\rangle=\sum_{N=1}^{\infty}\langle\psi,\psi_{N}\rangle\langle\psi_{N},\hat{H}\psi\rangle,
\]
the latter is true because $\hat{H}$ is a self-adjoint operator (and
we also know that $\psi\in\mathcal{D}(\hat{H})$). Thus, making use
of the closure relation,
\[
\langle\hat{H}\rangle_{\psi}=\sum_{N=1}^{\infty}\left[\int_{0}^{L}\mathrm{d}y\,\psi^{*}(y)\,\psi_{N}(y)\right]\left[\int_{0}^{L}\mathrm{d}z\,\psi_{N}^{*}(z)(\hat{H}\psi)(z)\right]
\]
\[
=\int_{0}^{L}\mathrm{d}y\,\int_{0}^{L}\mathrm{d}z\,\psi^{*}(y)\left[\sum_{N=1}^{\infty}\psi_{N}(y)\psi_{N}^{*}(z)\right](\hat{H}\psi)(z)=\int_{0}^{L}\mathrm{d}y\,\int_{0}^{L}\mathrm{d}z\,\psi^{*}(y)\,\delta(y-z)(\hat{H}\psi)(z).
\]
Finally,
\begin{equation}
\langle\hat{H}\rangle_{\psi}=\int_{0}^{L}\mathrm{d}y\,\psi^{*}(y)(\hat{H}\psi)(y)=\langle\psi,\hat{H}\psi\rangle.
\end{equation}
Note that the latter expression makes sense just because $\psi\in\mathcal{D}(\hat{H})$.
As shown in the Introduction, using the formula given in Eq. (37)
we obtain the result  $\langle\hat{H}\rangle_{\psi}=5\hbar^{2}/\mathrm{m}L^{2}$.
On the other hand, the expression given in Eq. (37) also implies the
expression given in Eq. (36). Certainly, we can prove this by using
the spectral decomposition of $\hat{H}$ and the action of the projector
operator $\hat{P}_{N}$, namely,
\[
\langle\hat{H}\rangle_{\psi}=\langle\psi,\hat{H}\psi\rangle=\left\langle \psi,\left(\sum_{N=1}^{\infty}E_{N}\hat{P}_{N}\right)\psi\right\rangle =\left\langle \psi,\sum_{N=1}^{\infty}E_{N}(\hat{P}_{N}\psi)\right\rangle =\left\langle \psi,\sum_{N=1}^{\infty}E_{N}\langle\psi_{N},\psi\rangle\psi_{N}\right\rangle 
\]
\begin{equation}
=\sum_{N=1}^{\infty}E_{N}\langle\psi_{N},\psi\rangle\langle\psi,\psi_{N}\rangle=\sum_{N=1}^{\infty}E_{N}\left|\langle\psi_{N},\psi\rangle\right|^{2},
\end{equation}
which is the result previously stated in Eq. (8). Consequently, the
mean value of the Hamiltonian operator given in Eqs. (1) and (2) when
it is in the state given in Eq. (4) can be calculated using Eq. (36)
or Eq. (37); in each case, the result is the same. 

Let us now consider the operator $\hat{H}^{2}=\hat{H}\hat{H}$ given
in Eq. (14), with its domain given in Eq. (16). Naturally, the mean
value of this operator in the state given in Eq. (4) can be written
as follows:
\begin{equation}
\langle\hat{H}^{2}\rangle_{\psi}=\sum_{N=1}^{\infty}E_{N}^{2}\left|\langle\psi_{N},\psi\rangle\right|^{2}.
\end{equation}
Certainly, the eigenvectors of $\hat{H}^{2}$ are the same as those
of $\hat{H}$, and their eigenvalues are the squares of those of $\hat{H}$
(obviously, the eigenfunctions of $\hat{H}^{2}$ belong to $\mathcal{D}(\hat{H}^{2})$).
Developing the right-hand side of the equation in (39), we obtain
the following results: 
\[
\langle\hat{H}^{2}\rangle_{\psi}=\sum_{N=1}^{\infty}E_{N}\langle\psi_{N},\psi\rangle E_{N}\langle\psi,\psi_{N}\rangle=\sum_{N=1}^{\infty}\langle E_{N}\psi_{N},\psi\rangle\langle\psi,E_{N}\psi_{N}\rangle=\sum_{N=1}^{\infty}\langle\hat{H}\psi_{N},\psi\rangle\langle\psi,\hat{H}\psi_{N}\rangle,
\]
the latter is because $E_{N}\in\mathbb{R}$. Thus,
\[
\langle\hat{H}^{2}\rangle_{\psi}=\sum_{N=1}^{\infty}\langle\psi_{N},\hat{H}\psi\rangle\langle\hat{H}\psi,\psi_{N}\rangle=\sum_{N=1}^{\infty}\langle\hat{H}\psi,\psi_{N}\rangle\langle\psi_{N},\hat{H}\psi\rangle.
\]
The latter is true because $\hat{H}$ is a self-adjoint operator (and
it can act perfectly on the state $\psi$). Thus,
\[
\langle\hat{H}^{2}\rangle_{\psi}=\sum_{N=1}^{\infty}\left[\int_{0}^{L}\mathrm{d}y\,(\hat{H}\psi)^{*}(y)\,\psi_{N}(y)\right]\left[\int_{0}^{L}\mathrm{d}z\,\psi_{N}^{*}(z)(\hat{H}\psi)(z)\right]
\]
\[
=\int_{0}^{L}\mathrm{d}y\,\int_{0}^{L}\mathrm{d}z\,(\hat{H}\psi)^{*}(y)\left[\sum_{N=1}^{\infty}\psi_{N}(y)\psi_{N}^{*}(z)\right](\hat{H}\psi)(z)
\]
\[
=\int_{0}^{L}\mathrm{d}y\,\int_{0}^{L}\mathrm{d}z\,(\hat{H}\psi)^{*}(y)\,\delta(y-z)(\hat{H}\psi)(z).
\]
Therefore, using the closure relation, we obtain the following result:
\begin{equation}
\langle\hat{H}^{2}\rangle_{\psi}=\int_{0}^{L}\mathrm{d}y\,(\hat{H}\psi)^{*}(y)(\hat{H}\psi)(y)=\langle\hat{H}\psi,\hat{H}\psi\rangle.
\end{equation}
Note that the mean value of the Hamiltonian squared is not equal to
$\langle\psi,\hat{H}\hat{H}\psi\rangle$. Actually, to write the relation
$\langle\hat{H}\psi,\hat{H}\psi\rangle=\langle\psi,\hat{H}\hat{H}\psi\rangle$,
the action of $\hat{H}$ on $\psi$ should be part of the domain of
$\hat{H}$, i.e., $\hat{H}\psi$ must satisfy the Dirichlet boundary
condition. As we know, this is not the case here (in fact, one has
that $\psi''(0)=\psi''(L)=-2\sqrt{30/L^{5}}$). Thus, the mean value
of the operator $\hat{H}^{2}$ in the state $\psi$ given in Eq. (4)
is not provided by its most common expression $\langle\psi,\hat{H}^{2}\psi\rangle$.
Substituting the state $\psi$ into Eq. (40), we find that $\langle\hat{H}^{2}\rangle_{\psi}=30\hbar^{4}/\mathrm{m}^{2}L^{4}$.
On the other hand, it was formally shown in the Introduction that
$\langle\psi,\hat{H}^{2}\psi\rangle=0$ (this is because $\hat{H}^{2}\psi=0$);
however, because $\psi$ does not belong to $\mathcal{D}(\hat{H}^{2})$
(i.e., because $\hat{H}\psi$ does not belong to $\mathcal{D}(\hat{H})$),
$\langle\psi,\hat{H}^{2}\psi\rangle$ is a futile quantity and cannot
represent the mean value of the operator $\hat{H}^{2}$ in the state
$\psi$. On the other hand, the result given in Eq. (40) also implies
the result given in Eq. (39). In effect, 
\[
\langle\hat{H}^{2}\rangle_{\psi}=\langle\hat{H}\psi,\hat{H}\psi\rangle=\left\langle \left(\sum_{N=1}^{\infty}E_{N}\hat{P}_{N}\right)\psi,\left(\sum_{M=1}^{\infty}E_{M}\hat{P}_{M}\right)\psi\right\rangle =\left\langle \sum_{N=1}^{\infty}E_{N}(\hat{P}_{N}\psi),\sum_{M=1}^{\infty}E_{M}(\hat{P}_{M}\psi)\right\rangle 
\]
\[
=\left\langle \sum_{N=1}^{\infty}E_{N}\langle\psi_{N},\psi\rangle\psi_{N},\sum_{M=1}^{\infty}E_{M}\langle\psi_{M},\psi\rangle\psi_{M}\right\rangle =\sum_{N=1}^{\infty}\sum_{M=1}^{\infty}E_{N}E_{M}\left\langle \,\langle\psi_{N},\psi\rangle\psi_{N},\langle\psi_{M},\psi\rangle\psi_{M}\,\right\rangle 
\]
\[
=\sum_{N=1}^{\infty}\sum_{M=1}^{\infty}E_{N}E_{M}\,\langle\psi_{N},\psi\rangle^{*}\,\langle\psi_{M},\psi\rangle\left\langle \psi_{N},\psi_{M}\right\rangle 
\]
\begin{equation}
=\sum_{N=1}^{\infty}\sum_{M=1}^{\infty}E_{N}E_{M}\,\langle\psi_{N},\psi\rangle^{*}\,\langle\psi_{M},\psi\rangle\,\delta_{M,N}=\sum_{N=1}^{\infty}E_{N}^{2}\left|\langle\psi_{N},\psi\rangle\right|^{2}.
\end{equation}
Naturally, the result obtained here is given by $\langle\hat{H}^{2}\rangle_{\psi}=30\hbar^{4}/\mathrm{m}^{2}L^{4}$.
Thus, one can conclude that both the first and the second equality
in Eq. (12) are meaningless (because $\psi$ does not belong to the
domain of $\hat{H}^{2}$). Obviously, the first equality in the second
line in Eq. (12) is true because, as we have just seen, the expression
obtained in Eq. (41) leads to the correct result. Finally, the root
mean square deviation of $\hat{H}$ in the state $\psi(x)$ given
in Eq. (4) is provided by
\begin{equation}
(\Delta\hat{H})_{\psi}=\sqrt{\langle\hat{H}^{2}\rangle_{\psi}-\langle\hat{H}\rangle_{\psi}^{2}}=\sqrt{\langle\hat{H}\psi,\hat{H}\psi\rangle-\langle\psi,\hat{H}\psi\rangle^{2}}=\frac{\sqrt{5}\hbar^{2}}{\mathrm{m}L^{2}},
\end{equation}
which is the result previously reported in Eq. (13).

\section{Final discussion}

\noindent Problems and contradictions such as the paradox we have
analyzed here become evident when the typical objects and elements
of standard wave mechanics are present (for example, the concrete
Hilbert space $\mathcal{L}^{2}(\Omega)$, where $\Omega$ could be
a finite interval, a semi-infinite interval, or an infinite interval
of the real line). Incidentally, this scheme has also been the most
widely used. Complications arise only when mathematical concepts are
manipulated without sufficient rigor (for example, if one does not
take into account the domains of definition of the unbounded operators
involved). On the other hand, it has been known for some time that
if one uses Dirac's ket and bra notation rigidly, problems also arise
(see Ref. \cite{RefA} and references therein), but these problems
are of a much more fundamental nature. For example, Dirac's symbolic
calculus does not allow us to give a precise meaning to the adjoint
of an unbounded operator \cite{RefA}. For this and other reasons,
we have not made use of the Dirac formalism in this paper, although
we recognize that the Dirac notation alone provides an elegant way
to write certain expressions in quantum mechanics. We think that the
work presented here will be of interest to all physicists and physics
students who are interested in the mathematics of quantum mechanics
and its correct teaching. 

\subsection*{Appendix}

\noindent As we have seen, the resolution of the paradox is intimately
related to the existence of the mean value of the operator $\hat{H}^{2}$.
We now show how this mean value can be obtained in an entirely formal
way, that is, without addressing the restrictions imposed by the domains
of the operators involved. Thus, in this section, we are not especially
concerned with the characteristics of the functions on which the operators
can act. 

As we have seen in the Introduction, the mean value of the Hamiltonian
operator $\hat{H}$ given in Eq. (1) in the state $\psi$ given in
Eq. (4) does not present any problem. The result is given in Eq. (6).
Now, we note that, after applying the method of integration by parts
twice, the following result is formally true: 
\[
\langle\hat{H}\psi,\hat{H}\psi\rangle=\langle\psi,\hat{H}^{2}\psi\rangle+\frac{\hbar^{4}}{4\mathrm{m^{2}}}\left.\left[\,\psi'^{*}(x)\,\psi''(x)-\psi^{*}(x)\,\psi'''(x)\,\right]\right|_{0}^{L}.\tag{A1}
\]
Assuming that the mean value of the operator $\hat{H}^{2}$ is obtained
from the formula given in Eq. (39), that is, $\langle\hat{H}^{2}\rangle_{\psi}=\langle\hat{H}\psi,\hat{H}\psi\rangle$
(i.e., simply consenting to the most natural definition of $\langle\hat{H}^{2}\rangle_{\psi}$),
and because $\hat{H}^{2}\psi=0$, from Eq. (A1), the following result
is obtained: 
\[
\langle\hat{H}^{2}\rangle_{\psi}=\frac{\hbar^{4}}{4\mathrm{m^{2}}}\left[\,\psi'^{*}(L)\,\psi''(L)-\psi'^{*}(0)\,\psi''(0)\,\right].\tag{A2}
\]
Again, we have that $\langle\hat{H}^{2}\rangle_{\psi}=30\hbar^{4}/\mathrm{m}^{2}L^{4}$.
Thus, in this case, the mean value is obtained by simply evaluating
a quantity at both ends of the box and then subtracting the two results.
Certainly, the boundaries of the interval in which the particle moves
and the boundary conditions imposed there play a definite role in
the final result. 

\section*{Conflicts of interest}

\noindent The authors declare no conflicts of interest.


\begin{thebibliography}{10}
\bibitem{RefA}F. Gieres, \textquotedblleft{}Mathematical surprises
and Dirac's formalism in quantum mechanics,\textquotedblright{} Rep.
Prog. Phys. \textbf{63}, 1893-931 (2000).

\bibitem{RefB}G. Bonneau, J. Faraut, G. Valent, \textquotedblleft{}Self-adjoint
extensions of operators and the teaching of quantum mechanics,\textquotedblright{}
Am. J. Phys. \textbf{69}, 322-31 (2001); Preprint, arXiv:0103153v1
{[}quant-ph{]} (2001). {[}The latter preprint is an extended version
of the published article that has various mathematical details.{]}

\bibitem[3]{RefC}D. Grau, \"{U}bungsaufgaben zur Quantentheorie\textemdash{}Quantentheoretische
Grundlagen, Version 5.52 {[}2020-06-20{]}, \copyright Dietrich Grau (2005).
To date, this is the last version of the book, which can be downloaded
from the author's website: http://dietrich-grau.at/download.html

\bibitem[4]{RefD}D. M. Gitman, I. V. Tyutin, B. L. Voronov, Self-adjoint
Extensions in Quantum Mechanics, Springer, New York (2012). 

\bibitem[5]{RefE}K. S. Ranade, ``Functional analysis and quantum
mechanics: an introduction for physicists,'' Fortschr. Phys. \textbf{63},
644\textendash{}658 (2015).

\bibitem[6]{RefF}D. H. T. Franco, L. S. Lima, ``A simple application
of the time evolution operator in the solution of a paradox in quantum
mechanics,'' Quantum Stud.: Math. Found. \textbf{5}, 273\textendash{}8
(2018).

\bibitem[7]{RefG}A. Cintio, A. Michelangeli, ``Self-adjointness
in quantum mechanics: a pedagogical path,'' Quantum Stud.: Math.
Found. \textbf{8}, 271\textendash{}306 (2021).

\bibitem[8]{RefH}C. Cohen-Tannoudji, B. Diu, F. Lalo\"{e}, Quantum Mechanics
-Volume I-, 2nd edn. Wiley-VCH, Weinheim (2020).

\bibitem[9]{RefI}T. F. Jordan, Linear Operators for Quantum Mechanics,
Dover Publications Inc, Mineola (2007).

\bibitem[10]{RefJ}M. Schechter, Operator Methods in Quantum Mechanics,
Elsevier North Holland Inc, New York (1981). 

\bibitem[11]{RefK}A. Messiah, Quantum Mechanics -Two volumes bound
as one-, Dover, Mineola (1999). 

\bibitem[12]{RefL}M. Amaku, F. A. B. Coutinho, O. J. P. \'{E}boli, E.
Massad, \textquotedblleft{}Some problems with the Dirac Delta function:
Divergent series in physics,\textquotedblright{} Braz. J. Phys. \textbf{51},
1324\textendash{}32 (2021).

\bibitem[13]{RefM}C. M. Bender, D. C. Brody, M. F. Parry, \textquotedblleft{}Making
sense of the divergent series for reconstructing a Hamiltonian from
its eigenstates and eigenvalues,\textquotedblright{} Am. J. Phys.
\textbf{88}, 148-52 (2020). \end{thebibliography}
\end{document}